# Sub-Doppler Laser Cooling of Fermionic $^{40}$K Atoms

G. Modugno*, C. Benkő†, P. Hannaford‡, G. Roati§, and M. Inguscio**

*INFM–European Laboratory for Non Linear Spectroscopy (L.E.N.S.), Università di Firenze, Largo E. Fermi, 2, I-50125 Firenze, Italy*

(July 9, 1999)

We report laser cooling of fermionic $^{40}$K atoms, with temperatures down to (15±5) $\mu$K, for an enriched sample trapped in a MOT and additionaly cooled in optical molasses. This temperature is a factor of 10 below the Doppler-cooling limit and corresponds to an rms velocity within a factor of two of the lowest realizable rms velocity ($\sim 3.5 v_{rec}$) in 3D optical molasses. Realization of such low atom temperatures, up to now only accessible with evaporative cooling techniques, is an important precursor to producing a degenerate Fermi gas of $^{40}$K atoms.

Ultracold dilute vapours of alkali atoms have proven to be favourable media for studying quantum degeneracy in bosonic atoms such as $^{87}$Rb, $^{7}$Li and $^{23}$Na [1–3]. One of the next major goals in this new field is to realize a degenerate Fermi gas of atoms, in which weakly interacting fermionic atoms are trapped in quantum degenerate states.

Among the various stable alkali isotopes, the only fermionic atoms are the weakly abundant isotopes $^{6}$Li (7.4%, I=1) and $^{40}$K (0.012%, I=4). In the case of $^{6}$Li, laser cooling is restricted by the large photon recoil velocity and the unresolved excited-state hyperfine splittings to temperatures of the order of 1 mK [4,5]. The bosonic isotopes of potassium, $^{39}$K and $^{41}$K, also have small excited-state ($4^2P_{3/2}$) hyperfine splittings, and laser cooling has thus far been restricted to temperatures close to the Doppler-cooling limit of 150 $\mu$K [6,7]. Fermionic $^{40}$K, on the other hand, has a moderately large and, more importantly, an inverted $4^2P_{3/2}$ hyperfine splitting (Fig.1), which should permit the use of large red detunings and hence might be expected to allow efficient sub-Doppler laser cooling, in much the same way as for $^{23}$Na.

Preliminary laser-cooling studies using a natural abundance sample of potassium gave an initial indication of temperatures down to about 50 $\mu$K [8], providing possible evidence of sub-Doppler cooling of $^{40}$K. Recent work [9] using an enriched sample of potassium (5% $^{40}$K) yielded laser-cooling temperatures of about 150 $\mu$K in a "Doppler-cooling stage" (in which the trapping light was switched closer to resonance). This was followed by an evaporative cooling stage that yielded temperatures down to 10 $\mu$K by s-wave collisions using mixed spin states and 5 $\mu$K by p-wave collisions using spin-polarized atoms.

In this paper we report the results of a systematic study of laser cooling of $^{40}$K atoms in optical molasses and demonstrate that temperatures down to 15 $\mu$K, i.e. about an order of magnitude below the Doppler limit, can be reached with pure optical cooling.

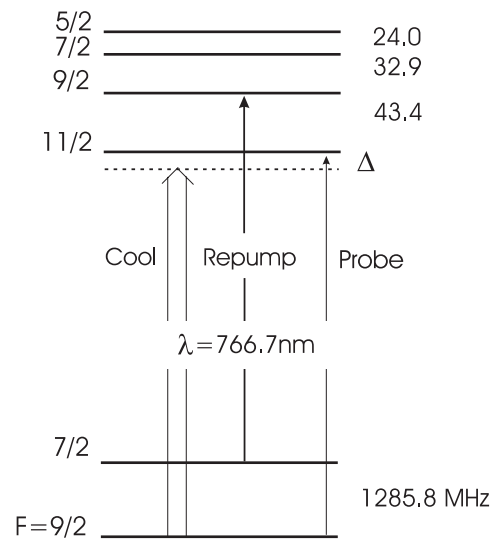

FIG. 1. Energy level scheme of the hyperfine structure of the D$_2$ transition of $^{40}$K. The cooling and repumping frequencies for magneto-optical trapping and cooling are generated by a single Ti:Sa laser, while the probe frequency is provided by a diode laser.

---

*modugno@lens.unifi.it

†permanent address: Janus Pannonius University, Pécs 7624, Hungary

‡permanent address: CSIRO Manufacturing Science and Technology, Clayton 3169, Australia

§also Dipartimento di Fisica, Università di Milano, Via Celoria 16, I-20100 Milano, Italy

**also Dipartimento di Fisica, Università di Firenze, Largo E. Fermi, 2, I-50125 Firenze, Italy

The experimental arrangement is basically the same as that described in our preliminary investigation [8], but with some improvements. We now use an enriched sample of $^{40}$K atoms (3% $^{40}$K), which permits up to $10^4$ times more atoms to be loaded into the magneto-optical trap; we use a home-made titanium:sapphire laser which is pumped by a low-noise 10 W doubled Nd:YVO$_4$ laser (Millenia X, Spectra Physics), and has the capability of yielding linewidths below 100 kHz; we determine the temperature of the atoms by a time-of-flight (TOF) absorption technique, which is much more accurate at low temperatures than the release-and-recapture method used previously.

Potassium-40 atoms are captured and cooled directly from an atomic vapour in a standard magneto-optical trap (MOT) using the F=9/2 → F'=11/2 cycling hyperfine transition of the 766.7 nm line in a standard six-beam $\sigma^+\sigma^-$ configuration. A repumping laser beam derived from the laser cooling beams is tuned to the F=7/2 → F'=9/2 hyperfine transition (see Fig.1). The atoms are then further cooled for 4-10 ms in an optical molasses stage with the magnetic field of the MOT switched off and the trapping laser switched to detunings up to $\Delta$=-7 Γ (where Γ=2π×6.2 MHz) and intensities down to about 3 $I_{sat}$ (where $I_{sat}$=1.8 mW cm$^{-2}$).

positioned 15 mm beneath the MOT. The atom temperatures are determined by fitting theoretical curves to the TOF absorption data.

A time-of-flight absorption signal for atoms released directly from the MOT is shown in Fig.3(a). For a typical laser-atom detuning of $\Delta$=-3 Γ and trapping laser intensity of about 17 $I_{sat}$, which are required to optimize the loading process, the temperature of potassium atoms in the MOT is already down to (87±5) μK, prior to cooling in optical molasses. The temperature can be further reduced to approximately 60 μK, by lowering the laser intensity. This result confirms our preliminary estimate of the $^{40}$K atom temperature in a MOT [8], which was determined using the release-and-recapture method.

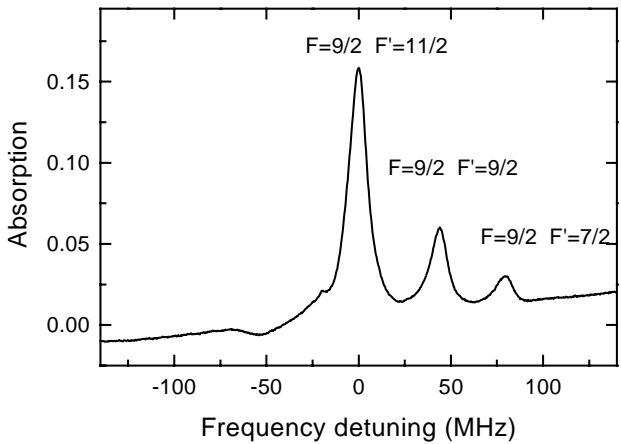

FIG. 2. Frequency scan of the atomic absorbtion due to the F=9/2 → F'=11/2, 9/2, 7/2 hyperfine triplet from a sample of laser-cooled $^{40}$K atoms in the MOT.

The total number of $^{40}$K atoms in the MOT is estimated from measurements of the intensity of light emitted by the atoms. The number density of atoms in the MOT is also determined, by measuring the atomic absorption of a weak, narrow-band diode-laser probe beam tuned to the F=9/2 → F'=11/2 transition. From a knowledge of the size of the trap, cross checks are made against the total number of atoms in the MOT. A frequency scan of the atomic absorption signal for the F=9/2 → F'=11/2, 9/2, 7/2 hyperfine triplet from a sample of laser-cooled K atoms in the MOT is shown in Fig.2. With the trapping laser beams detuned $\Delta$=-3 Γ to the red of resonance, up to $10^8$ $^{40}$K atoms are estimated to be captured in a volume of diameter about 1.5 mm. This corresponds to a density of $^{40}$K atoms of about $10^{10}$ cm$^{-3}$. The MOT loading time is around 3 s. Time-of-flight (TOF) absorption signals are recorded using a weak, thin (∼0.5 mm) sheet of narrow-band, diode-laser light tuned to the F=9/2 → F'=11/2 transition and

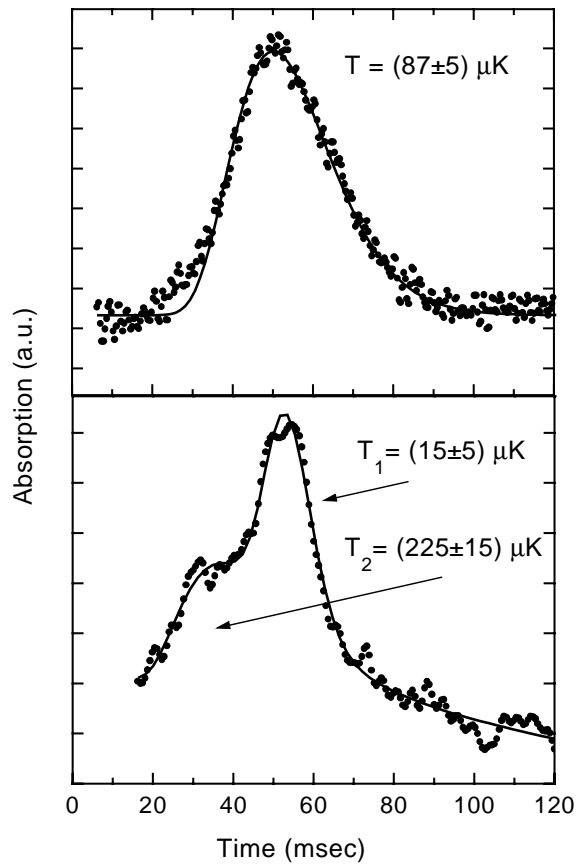

FIG. 3. Time-of-flight signals from cold $^{40}$K atoms, as detected in absorption by a weak sheet of light. In (a) the atoms are falling directly from the MOT, loaded with a trapping laser detuning $\Delta$=-3 Γ and intensity of 17 $I_{sat}$. In (b), the atoms have experienced an additional molasses phase (4 ms) with parameters $\Delta$=-6.6 Γ and I=4 $I_{sat}$; two different velocity classes are present, yielding different temperatures.

To achieve even lower temperatures, we have explored the effect of additional cooling in an optical molasses stage with the magnetic field of the MOT switched off and the trapping laser suddenly switched to larger laser-atom detunings and lower laser intensities. Fig.3(b) shows the TOF absorption signal recorded after cooling the atoms for 4 ms in optical molasses at a detuning of $\Delta$=-6.6 $\Gamma$ and laser intensity of about 4 $I_{sat}$. We notice the presence of two distinct velocity classes in the atomic cloud: a fit to the narrow component yields an atom temperature of (15±5) $\mu$K, while the broad component corresponds to (225±15) $\mu$K. The TOF signal shown in Fig.3(b) represents the sum of 10 consecutive recordings, and the uncertainty in the temperature is due mainly to the accuracy of the fitting procedure. We estimate possible corrections for the trap size and probe beam thickness to be negligibly small.

The temperature of the narrow velocity component ((15±5) $\mu$K) is a factor of ten below the Doppler-cooling limit for $^{40}$K and corresponds to an rms atom velocity of 9.7 cm s$^{-1}$, or 7.5 times the single-photon recoil velocity, $v_{rec}$=1.3 cm s$^{-1}$. This is within a factor of about two of the lowest realizable rms velocity achievable in 3D optical molasses [10–12]. We speculate that the broad component can arise from K atoms that aren't sufficiently cold at the time of switching from the MOT to the optical molasses to experience sub-Doppler forces; i.e. the velocities of the atoms at the end of the MOT stage are outside the sub-Doppler velocity capture range as determined by the values of the laser-atom detuning and laser intensity during the molasses stage. As a result, this class of atoms can experience only Doppler forces, which lead to even higher temperatures than those achieved in the MOT. Similar bimodal velocity distributions have previously been observed in early experiments on atomic sodium [13]. As confirmation of this possible explanation, the broad component was observed only for large values of the laser-atom detuning (-5 $\Gamma$ to -7 $\Gamma$), while for small detunings only the cold component was discernable. In principle, it should be possible to reduce the broad background component by adiabatically changing the detuning and the intensity of the cooling laser at the start of the optical molasses stage.

As usually found in optical molasses, the cooling efficiency was strongly dependent on a good balancing of the laser beams, and in nulling any residual magnetic field. Fig.4 shows a typical dependence of the atom temperature on laser-atom detuning for a range of detunings. As the detuning is increased from $\Delta$=-2 $\Gamma$ to -6 $\Gamma$, the temperature of the atoms is reduced from about 70 $\mu$K to about 20 $\mu$K. A similar behavior is observed when the laser intensity is reduced from about 15 $I_{sat}$ down to about 3 $I_{sat}$. The number of cold atoms in the optical molasses, as determined from the area under the TOF signal, is found to decrease with the atom temperature when reducing the laser intensity or increasing the laser-atom detuning. We have observed a drop by almost an order of magnitude when the temperature is lowered from 70 $\mu$K to about 20 $\mu$K, and for intensities lower than 3 $I_{sat}$ and detunings larger than -7 $\Gamma$ we were unable to distinguish the TOF signal from the noise. As mentioned in the earlier discussion, we anticipate we should be able to reduce the loss of cold atoms by adiabatically changing from the MOT to the optical molasses.

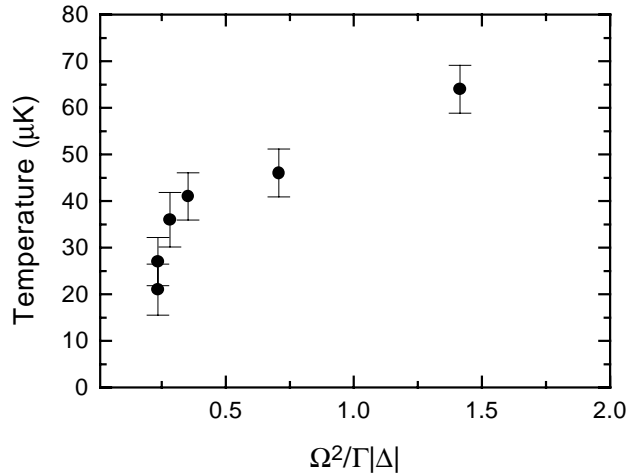

FIG. 4. Typical behaviour of the atomic temperature as a function of the detuning of the cooling molasses beams, for I=3 $I_{sat}$ and $\Delta$=-2 to -6 $\Gamma$, for a molasses phase lasting 5 ms.

In summary, sub-Doppler laser cooling, with temperatures down to about 15 $\mu$K, have been realized for fermionic $^{40}$K atoms cooled in optical molasses. This temperature is a factor of 10 below the Doppler-cooling limit and corresponds to an rms velocity within a factor of two of the lowest realizable rms velocity ($\sim$3.5 $v_{rec}$) in 3D optical molasses. We note that up to now the temperature range we have accessed with laser cooling had been reached only by means of more complex evaporative cooling techniques [9]. On the other hand, these temperatures should provide an excellent starting point for implementing evaporative cooling to obtain the very low temperatures likely to be required for producing a degenerate Fermi gas of $^{40}$K atoms. Alternatively, such temperatures should also provide a suitable environment for confining $^{40}$K atoms in an optical dipole trap for further studies, such as those on collisional properties at low temperatures. Eventually, even in such trap one could exploit additional cooling methods to increase the phase-space density towards degeneracy.

Although the observed temperatures are close to the limit imposed by the photon recoil, we anticipate that in future it should be possible to realize even lower temperatures for $^{40}$K by improving the efficiency of polarization gradient cooling, for example, by further optimization of the laser-atom detuning, laser intensity and stray magnetic fields; by adiabatically changing the detuning and intensity of the cooling laser at the start of the optical molasses stage; and by using optical molasses with beams having linear $\pi^x$ $\pi^y$ polarizations instead of $\sigma^+\sigma^-$ polarizations [12]. Such improvements could also prove to be useful in increasing the efficiency of the molasses in cooling a large number of atoms.

Besides being able to be cooled to low temperatures, fermionic K has a number of other characteristics that makes it an attractive candidate for studying quantum degeneracy. Potassium-40 atoms in the $|F=9/2,m=9/2\rangle$ state are known to have a large positive triplet scattering length, both for p-wave elastic scattering between like spin states and for s-wave elastic scattering between mixed spin states [9,14]. Thus p-wave scattering, whose cross section scales quadratically with energy, should allow evaporative cooling to proceed at temperatures down to about 20 $\mu$K [9], while s-wave scattering between mixed-spin states should allow evaporative cooling to the very low temperatures ($\sim$1 $\mu$K [15]) likely to be required to reach quantum degeneracy. Also, owing to the inverted ground hyperfine structure in $^{40}$K (Fig.1), certain spin-exchange collisions, for example, between the $|F=9/2,m=9/2\rangle$ and $|F=9/2,m=7/2\rangle$ states and also hyperfine-changing collisions, are energetically forbidden at low temperatures [14]. Finally, the ground hyperfine states F=9/2, 7/2 of $^{40}$K have rich Zeeman structure, with nine low-field seeking spin states (in the high-field I-J decoupled limit) that are magnetically trappable [14].

We aknowledge useful suggestions and technical help from F.S. Cataliotti, C. Fort, J. Dalibard, F. Minardi, and M. Prevedelli. The Ti:Sa laser was constructed according to a scheme by F. Biraben, with the help of F. Marin and F. Nez. This work was supported by the European Community Council (ECC) under Contract No. ERBFMGECT950017, by INFM "Progetto di Ricerca Avanzata" and by CNR "Progetto integrato" Contract No. 96.00068.02. C.B. was founded by ECC in the frame of Socrates/Erasmus Project; P.H. would like to thank LENS for financial support and M. Inguscio for his kind and generous hospitality during a recent visit to LENS; G.R. would like to thank L. Reatto for his encouragement.